\documentclass[12pt, letterpaper]{article}
\usepackage{jheppub}
\usepackage{amsmath}
\usepackage{amssymb}
\usepackage{epsfig}
\usepackage{graphicx}
\usepackage{hyperref}
\usepackage{tikz}
\usetikzlibrary{decorations.pathmorphing}
\usepackage{pgfplots}
\pgfplotsset{width=6cm}
\usepackage{graphicx}

\newcommand{\comment}[1]{}
\newcommand{\oprod}[2]{\left| #1 \right\rangle\!\! \left\langle #2 \right| } 
\newcommand{\ket}[1]{\left|#1\right>}
\newcommand{\ev}[1]{\left\langle#1\right\rangle}

\newcommand{\be}{\begin{equation}}
\newcommand{\ee}{\end{equation}}

\title{Entanglement Tsunami in (1+1)-Dimensions}
\author{Stefan Leichenauer}
\affiliation{Center for Theoretical Physics and Department of Physics,\\
 University of California, Berkeley, CA 94720, U.S.A.}
\emailAdd{sleichen@berkeley.edu}

\author{and Mudassir Moosa}
\emailAdd{mudassir.moosa@berkeley.edu}

\abstract{We study the time dependence of the entanglement entropy of disjoint intervals following a global quantum quench in (1+1)-dimensional CFTs at large-$c$ with a sparse spectrum. The result agrees with a holographic calculation but differs from the free field theory answer. In particular, a simple model of free quasiparticle propagation is not adequate for CFTs with a holographic dual. We elaborate on the entanglement tsunami proposal of Liu and Suh and show how it can be used to reproduce the holographic answer.}

\begin{document}
\maketitle

\section{Introduction}

Computation of entanglement entropy in quantum field theory is a topic of growing interest. Unfortunately, in many situations it is an extremely difficult quantity to compute. For strongly coupled systems, one of our best analytical tools is holography. In a holographic theory, entanglement entropy is evaluated as the area of an extremal surface in the dual geometry \cite{RT,HRT}. In some cases, especially in (1+1)-dimensional CFTs, purely field-theoretical arguments can be given for the form of the entanglement entropy. 

We will be using entanglement entropy to study the approach to equilibrium following a global quantum quench. Calabrese and Cardy \cite{CC-particle,CC-particle-two} famously showed that, in (1+1)-dimensions, the entanglement entropy of an interval increases linearly with time until it saturates at the thermal value. The field-theoretical results could be reproduced by a simple intuitive model: the growth and saturation of the entanglement entropy was effectively modeled by the free propagation of quasiparticle excitations, which we will review below.

This linear growth in (1+1)-dimensions was confirmed in holographic calculations, first numerically \cite{Abajo}, and then analytically \cite{VB-one, VB-two}, where it was also extended to higher dimensions (see also \cite{Maldacena-Hartman}). In \cite{HL-one, HL-two}, it was noted that a free-streaming quasiparticle picture was inadequate to explain the rate of growth beyond the (1+1)-dimensional case and it was suggested that the linear growth could be thought of in terms of an ``entanglement tsunami," represented by an effective wavefront which propagates into the region under consideration and entangles interior degrees of freedom with exterior degrees of freedom. An underlying free-streaming quasiparticle model for the entanglement tsunami yielded a wavefront velocity smaller than the velocity calculated holographically in dimensions higher than (1+1). In (1+1) dimensions, the holographic calculation and quasiparticle model both said that the effective wavefront moved at the speed of light. At this level, then, it seemed that quasiparticles were sufficient to explain the time dependence of the entanglement entropy of a single interval in (1+1) dimensions, even for holographic CFTs which are strongly interacting.

However, when multiple disjoint intervals are considered, there is a qualitative difference between the holographic calculation and the quasiparticle model of Calabrese and Cardy (as noted in \cite{VB-three, Asplund}, and also present in \cite{Allais:2011ys}): the holographic calculation gives a non-decreasing entropy, while the quasiparticle model has both increasing and decreasing behavior as a function of time. Therefore already in (1+1)-dimensions there is a need to replace the quasiparticle picture. We will elaborate on the entanglement tsunami proposal, showing how it can be used as a rule to calculate the entanglement entropy of one or two intervals in holographic CFTs and provide a natural upper bound for more than two intervals. We will make no attempt to derive the entanglement tsunami from underlying physical excitations, quasiparticle-like or otherwise. 

The remainder of this paper is organized as follows. In Section~\ref{sec-review} we will review the setup for the problem and the results from both the quasiparticle and holographic viewpoints. In Section~\ref{sec-largec} we will provide a CFT argument for the holographic result, showing how CFTs with large-$c$ and a sparse spectrum differ from weakly coupled CFTs, and in particular do not follow the quasiparticle prediction. In Section~\ref{sec-tsunami} we will propose an entanglement tsunami prescription for the entanglement entropy as a function of time. Finally, in Section~\ref{sec-disc} we will describe how these results may be extended to higher dimensions, where there are still some unresolved issues, and speculate on possible derivations of the entanglement tsunami from interactions.

\section{Setup and Review}\label{sec-review}

\subsection{Quasiparticle Model}

We are interested in the entanglement entropy of a subregion $A$ of the real line following a global quench in a CFT. Note that $A$ is always the union of a collection of disjoint intervals. Let $A^c$ denote the complement of $A$. Globally, the system is in a pure state $\ket{\Psi}$, and the entanglement entropy for the region $A$ as given by
\be
S = -{\rm Tr}\, \rho_A \log \rho_A,\,\,\,\,\,\,\,\, \text{with }\rho_A = {\rm Tr}_{A^c} \oprod{\Psi}{\Psi}.
\ee
$\ket{\Psi}$ is a time-dependent state, and so $S$ will be time-dependent.  A global quench is defined by beginning in the vacuum state of one theory, and then suddenly changing the Hamiltonian to that of  a different theory. The result is that the initial state from the point of view of the second theory (the CFT) is a highly excited state, but has a simple entanglement structure. We are interested in the time dependence of the finite part of the entanglement entropy, so we will assume that the UV-divergent parts can be subtracted in a consistent way, whether or not they are modified by the quench.  Henceforth when we refer to the entropy we will mean the finite part of the entropy, or the vacuum-subtracted entropy. At $t=0$, then, we have $S=0$.

At late times, the system will effectively thermalize and we should find $S(t\to\infty) = S_{\rm therm} = s_{\rm eq} {\rm Vol}(A)$. In (1+1)-dimensions ${\rm Vol}(A)$ is just the sum of the lengths of the intervals that make up $A$, but we are emphasizing that the thermal entropy is extensive with the volume of the system. $s_{\rm eq}$ is the thermal entropy density, which is a property of the state. At intermediate times, either a CFT calculation or a holographic calculation can be used to describe the transition from zero entropy to the thermal result. We will discuss both of these calculations below, but for now we will record the expected answer as predicted by the quasiparticle model.

Calabrese and Cardy showed \cite{CC-particle,CC-particle-two} that, following a global quench, the time-dependence of the entropy can be effectively modeled by the propagation of entangled quasiparticles, at least for weakly coupled CFTs. At the time of the quench, we imagine that a uniform density of EPR pairs of quasiparticles are produced, where each pair begins localized at a point and consists of a left-mover and right-mover. These quasiparticles move in opposite directions at the speed of light, and there are no interactions between pairs. To compute the entanglement entropy of $A$, we only have to count the number of unpaired particles in the region at any given time: 
\be
S(t) \propto \int_{x' \in A}dx'\int_{x'' \in A^c }dx''\int_{-\infty}^{\infty}dx\,\Big\{\delta(x'-x-t)\delta(x''-x+t) +\delta(x'-x+t)\delta(x''-x-t)\Big\}.
\ee	
The constant of proportionality is related to the initial density of EPR pairs, which determines $s_{\rm eq}$. When $A$ consists of a single interval of length $L$, computing the above integral gives the following time dependence:
\begin{gather}
S(t)= 2s_{\rm eq} \times \left \{ \begin{array}{ll}
t, &~~~ t \le \frac{L}{2}, \\
\\
\frac{L}{2}, &~~~t>  \frac{L}{2}. \\
\end{array}    \label{qp-one}
\right.
\end{gather}\\
First there is a linear growth phase, and then saturation at the thermal value.

	\begin{figure}
	\centering 
\begin{tikzpicture}
\begin{axis}[
title = {Quasiparticle prediction},
axis lines = left,
xlabel = $2t$,
ylabel = {$S(t)$},
ytick={ 0},
xtick={0 },
ymin=0,
ymax=5.5,
]

\addplot[
domain=0:5, samples = 100
]
{x};

\addplot[
domain=5:10, samples = 100
]
{5};

\addplot[
domain=10:15, samples = 100
]
{10 - 0.5*x};

\addplot[
domain=15:20, samples = 100
]
{0.5*x - 5};

\addplot[
domain=20:30, samples = 100
]
{5};
\end{axis}

\draw[dotted] (0.75,0) --(0.75,3.6);
\node at (.7,-0.20) {\footnotesize$L$};

\draw[dotted] (1.46,0) --(1.46,3.6);
\node at (1.4,-0.20) {\footnotesize$R$};

\draw[dotted] (2.2,0) --(2.2,3.6);
\node at (2.2,-0.20) {\footnotesize$L+R$};

\draw[dotted] (2.95,0) --(2.95,3.6);
\node at (3.4,-0.20) {\footnotesize$2L+R$};


		\end{tikzpicture}%
			%
			%
\begin{tikzpicture}
\begin{axis}[
title = {Holographic prediction},
axis lines = left,
xlabel = $2t$,
ylabel = {$S(t)$},
ytick={0},
xtick={0},
ymin=0,
ymax=5.5,
]

\addplot[
domain=0:5, samples = 100
]
{x};

\addplot[
domain=5:30, samples = 100
]
{5};

\end{axis}

\draw[dotted] (0.75,0) --(0.75,3.6);
\node at (.75,-0.20) {\footnotesize$L$};



		\end{tikzpicture}

			
\caption{The entropy production as a function of time for a region consisting of two disjoint intervals of length $L$, separated by a distance $R>L$. The quasiparticle model (left) shows decreasing behavior between $2t=R$ and $2t=L+R$. The holographic calculation (right) is monotonically increasing before saturation at $2t=L$, after which the entropy remains constant.}	\label{s}		
		\end{figure}
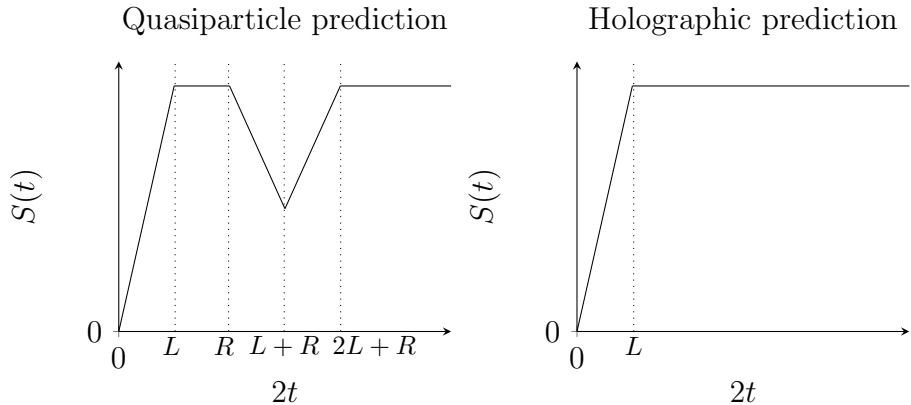

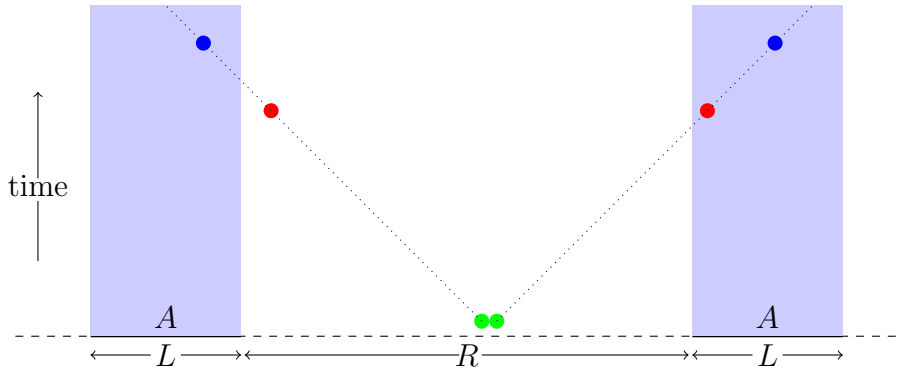
\begin{figure}
		\centering 
\begin{tikzpicture}
	\draw [dashed](0,0) --(12,0);
	\draw [thick](1,0) --(3,0);
	\draw [thick](9,0) --(11,0);
	\draw [->] (2.2,-.25) -- (3.0,-.25);
	\draw [->] (10.2,-.25) -- (11.0,-.25);
	\draw [->] (1.8,-.25) -- (1,-.25);
	\draw [->] (9.8,-.25) -- (9,-.25);
	\draw [->] (5.9,-.25) -- (3.05,-.25);
	\draw [->] (6.2,-.25) -- (8.95,-.25);
	\fill[blue!20!white] (1,0) rectangle (3,4.4);
	\fill[blue!20!white] (9,0) rectangle (11,4.4);

	\fill [green](6.2,.2) circle (.1cm);
	\fill [green](6.4,.2) circle (.1cm);
	
	\fill [red](3.4,3.0) circle (.1cm);
	\fill [red](9.2,3.0) circle (.1cm);

	\fill [blue](2.5,3.9) circle (.1cm);
	\fill [blue](10.1,3.9) circle (.1cm);

	\draw [dotted](6.2,0.2) --(2,4.4);
	\draw [dotted](6.4,0.2) --(10.6,4.4);
	
	\draw [->] (.3,2.1) -- (.3,3.25);
	\draw  (.3,1) -- (.3,1.8);
	\node at (.3,2) {time};
		\node at (2,-0.25) {$L$};
		\node at (10,-0.25) {$L$};
		\node at (6,-0.25) {$R$};
		\node at (2,+0.25) {$A$};
		\node at (10,0.25) {$A$};
		
 \end{tikzpicture}
\caption{An EPR pair produced at the points marked as green at the bottom of the figure. When the constituent particles are at the positions marked as red at the intermediate time, they contribute to the entanglement entropy. At the later time when the particles are at the positions marked as blue, they do not contribute to the entanglement entropy. This process leads to a decrease in the entanglement entropy in the quasiparticle picture.}
			\label{drop_in_the_entanglement_entropy}
		\end{figure}

The behavior is a little more complicated when $A$ consists of two disjoint intervals. For simplicity, consider the case where the two intervals have equal lengths $L$ and are separated by a distance $R>L$. Then the quasiparticle model gives (see Fig.~\ref{s} for a plot)
\begin{gather}
S(t)= 2s_{\rm eq} \times \left \{ \begin{array}{ll}
2t, &~~~ t \le \frac{L}{2}, \\
L, &~~~ \frac{L}{2}<t<\frac{R}{2}, \\
L - \Big(t - \frac{R}{2}\Big), &~~~ \frac{R}{2}<t<\frac{L+R}{2}, \\
L +\Big(t - L -\frac{R}{2}\Big), &~~~ \frac{L+R}{2}<t<\frac{2L+R}{2},\\
L, &~~~ t>\frac{2L+R}{2}.
\end{array}    
\right.
\end{gather}
In particular, this result tells us that the entanglement entropy is not monotonic in time. We can understand the drop in the entanglement entropy in the following way. Consider an EPR pair that started propagating from the region between the two intervals which make up $A$ at the time of quench, $t=0$ (see Fig.~\ref{drop_in_the_entanglement_entropy}). At some time $t<R/2$, one of the particles will enter region $A$ and while the other remains in the complement. At that time, and for times immediately following, this pair contributes one unit to the entanglement entropy. However, at a later time the second particle will enter the region $A$ as well, and so there is an opportunity for both particles to be inside $A$ at the same time. Then they will no longer contribute to the entanglement entropy. This results in decreased entanglement.

\subsection{Holographic Calculation}

The time evolution of entanglement entropy after a global quench can also be studied using the AdS-CFT correspondence  \cite{Abajo, VB-one, VB-two, HL-one, HL-two}. According to the AdS-CFT dictionary, a global quench in the boundary theory is dual to throwing a spatially homogenous and isotropic shell of matter into the bulk. This shell will eventually collapse to form a black hole, which is the gravity dual to a thermal state in the CFT.
		
\begin{figure}
\centering 
\begin{tikzpicture}
	\draw [red](0,-1.5) --(0,4);
	\draw [black](0,-1.5) --(-4.75,+3.25);
	\draw [black](-4,4) --(-4.75,3.25);
	\draw [green](0,0) --(-4,4);
	\draw [decorate,decoration=zigzag](-4,4) --(0,4);
	\draw [blue](0,4) --(-2,2);
		\node at (-5,3.25) {$z_{\infty}$};
		\draw   [dashed] (0,3.8) to[out=180,in=0] (-0.3,3.7);
		\draw   [dashed] (-0.3,3.7) to[out=180,in=0] (-3,2.5);
		\draw   [purple] (-0.2,4) to[out=190,in=0] (-3,2.55);
 \end{tikzpicture}
\caption{Penrose diagram of the time dependent geometry following the quench. The red vertical line on the right is the AdS boundary ($z=0$ in Poincare patch). The green diagonal line is the infalling shell, and the blue diagonal line is the horizon. The dashed curve is a late time extremal surface, which asymptotes to the critical surface, indicated by the solid curve. The linear growth of entanglement entropy comes from the portion of the extremal surface lying along the critical surface behind the horizon.}
			\label{penrose diagram}
		\end{figure}
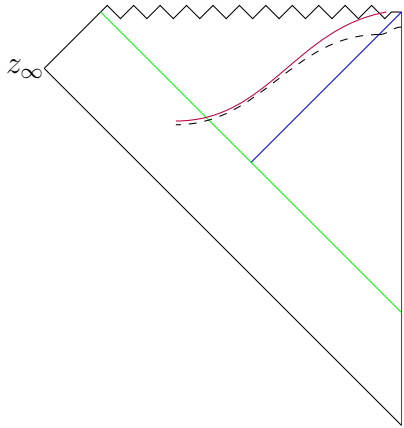

After the quench, the geometry of the bulk is given by the time-dependent AdS-Vaidya geometry (displayed as a conformal diagram in Fig.~\ref{penrose diagram}). As a result, the area of the boundary-anchored extremal surfaces, and hence the entanglement entropy of the boundary region, will depend on time. Though we are concerned with a (1+1)-dimensional CFT, and hence a 2+1-dimensional bulk, Liu and Suh were able to perform the calculation in arbitrary dimensions and with different types of black hole. 

The local equilibrium length is given by the horizon radius, $z_{h}$. Consider a spatial interval on the boundary of length $L \gg z_h$. For times $t \gg z_h$, but smaller than $L/2$, the extremal surface in the bulk anchored to the endpoints of the interval has an area that grows linearly with time. Geometrically, the linear growth is tied to the existence of a critical extremal surface behind the black hole horizon. The extremal surface anchored to the boundary interval goes behind the horizon and approaches the critical surface. The length of the portion of the extremal surface lying along the critical surface increases linearly with $t$, which leads to a linear growth in area. At $t \approx L/2$, there is a transition (the details of which do not concern us here), after which the extremal surface lies outside of the horizon in the black hole portion of the geometry and is no longer influenced by the collapsing shell. The symmetries of this geometry ensure that the area is time-independent in this region, and so this represents thermal saturation of the entropy.

To summarize, the vacuum subtracted entropy for a single interval as computed holographically is given by
\begin{gather}\label{eq-singleintervalholo}
S(t)=2s_{\text{eq}} \times \left \{ \begin{array}{ll}
t, &~~~ t \le \frac{L}{2}, \\
\frac{L}{2}, &~~~ t>  \frac{L}{2}. \\
\end{array}    
\right.
\end{gather}
where $s_{\text{eq}}$ is related to the AdS-radius, $L_{\text{AdS}}$, horizon radius, $z_{h}$, and Newton's constant, $G_{N}$, by
\begin{equation}
s_{\text{eq}} = \frac{1}{4G_{N}}\frac{2L_{\text{AdS}}}{z_{h}}.
\end{equation}
This holographic result for a single interval agrees with the predictions of the quasiparticle picture \eqref{qp-one}.

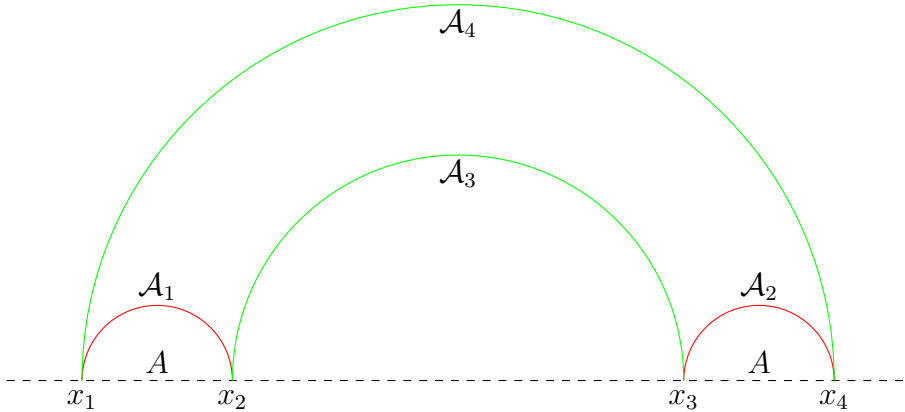
\begin{figure}
		\centering 
\begin{tikzpicture}
	\draw [dashed](0,0) --(12,0);
	\draw [red](3,0) arc[radius = 1, start angle = 0, end angle = 180];
	\draw [red](11,0) arc[radius = 1, start angle = 0, end angle = 180];
	\draw [green](9,0) arc[radius = 3, start angle = 0, end angle = 180];
	\draw [green](11,0) arc[radius = 5, start angle = 0, end angle = 180];
		\node at (1,-0.25) {$x_{1}$};
		\node at (3,-0.25) {$x_{2}$};
		\node at (9,-0.25) {$x_{3}$};
		\node at (11,-0.25) {$x_{4}$};
		\node at (2,1.25) {$\mathcal{A}_{1}$};
		\node at (10,1.25) {$\mathcal{A}_{2}$};
		\node at (6,2.75) {$\mathcal{A}_{3}$};
		\node at (6,4.75) {$\mathcal{A}_{4}$};
		\node at (2,+0.25) {$A$};
		\node at (10,0.25) {$A$};
 \end{tikzpicture}
\caption{Here we display the extremal surfaces for two intervals. The first candidate HRT surface is the union of the two smaller arcs (marked in red and labeled $\mathcal{A}_1$ and $\mathcal{A}_2$). The second candidate is the union of the two larger arcs (marked in green and labeled $\mathcal{A}_3$ and $\mathcal{A}_4$).}
			\label{HRT surfaces}
		\end{figure}

We can deduce the holographic result for an arbitrary collection of intervals by using the answer for a single interval. For example, consider the case of a pair of intervals, $[x_1,x_2]$ and $[x_3,x_4]$. We need to find the bulk extremal surface with minimal area anchored to those intervals on the boundary. This is called the HRT surface \cite{HRT}. There are two candidate HRT surfaces, which we display in Fig.~\ref{HRT surfaces}. First, there is the union of the bulk extremal surfaces associated to the two intervals  $[x_1,x_2]$ and $[x_3,x_4]$ individually  ($\mathcal{A}_1$ and $\mathcal{A}_2$ in Fig.~\ref{HRT surfaces}). But the union of the bulk extremal surfaces associated to  $[x_1,x_4]$ and $[x_2,x_3]$ is a second choice ($\mathcal{A}_3$ and $\mathcal{A}_4$ in Fig.~\ref{HRT surfaces}). The holographic prescription is to compute the total area in both cases and take the minimum value. But in each case the extremal surfaces are just unions of extremal surfaces associated to intervals, and the time-dependence of those surfaces is given by \eqref{eq-singleintervalholo}.

Let $L_1 = x_2-x_1$, $L_2 = x_4-x_3$, and $R = x_3-x_2$. Suppose $L_1 < L_2$. Then the first possible pair of extremal surfaces ($\mathcal{A}_1$ and $\mathcal{A}_2$ in Fig.~\ref{HRT surfaces}) would have a time-dependence given by
\be
S^{(1)}(t) = 2 s_{\rm eq} \times \begin{cases}
2t,~&t<\frac{L_1}{2},\\
t + \frac{L_1}{2},~&\frac{L_1}{2}<t<\frac{L_2}{2},\\
\frac{L_1+L_2}{2},~&t>\frac{L_2}{2},
\end{cases}
\ee
while the second choice of extremal surfaces ($\mathcal{A}_3$ and $\mathcal{A}_4$ in Fig.~\ref{HRT surfaces}) has
\be
S^{(2)}(t) = 2 s_{\rm eq} \times \begin{cases}
2t,~&t<\frac{R}{2},\\
t+\frac{R}{2},~&\frac{R}{2}<t<\frac{L_1+L_2+R}{2},\\
\frac{L_1+L_2+2R}{2},~&t>\frac{L_1+L_2+R}{2}.
\end{cases}
\ee
At each time we take the minimum of $S^{(1)}(t)$ and $S^{(2)}(t)$ to get $S(t)$. The interesting case is when $R$ is the smallest of the three lengths, which means that our two disjoint intervals are close together. Then we have
\be
S(t) = 2 s_{\rm eq} \times \begin{cases}
2t,~&t<\frac{R}{2},\\
t +\frac{R}{2},~&\frac{R}{2}<t<\frac{L_1+L_2-R}{2},\\
\frac{L_1+L_2}{2},~&t>\frac{L_1+L_2-R}{2}.
\end{cases}
\ee
In the other two cases, $L_1 < R < L_2$ and $L_1  < L_2<R$ , we have
\be
S(t) =2 s_{\rm eq} \times \begin{cases}
t,~&t<\frac{L_1}{2},\\
t +\frac{L_1}{2},~&\frac{L_1}{2}<t<\frac{L_2}{2},\\
\frac{L_1+L_2}{2},~&t>\frac{L_2}{2}.
\end{cases}
\ee
A plot of the case $L_1=L_2 < R$ is in Fig.~\ref{s}. Unlike the quasiparticle model, the holographic calculation gives a non-decreasing answer for the entropy. It is easy to see why this is the case. For an arbitrary boundary region, each of the candidate HRT surfaces is the union of a collection of extremal surfaces anchored on boundary intervals. But each surface anchored on a boundary interval has a non-decreasing area. Therefore each candidate HRT surface has a non-decreasing area, and so the true HRT surface has a non-decreasing area (even though the identity of the true surface may change as a function of time).

\section{Large-$c$ CFT Calculation}\label{sec-largec}

\subsection{Vacuum State Entanglement}

In this section we review the standard machinery for calculating the entanglement entropy of disjoint intervals in a (1+1)-dimensional CFT. The idea is to use the replica trick \cite{CC-particle-two,replica-one,replica-two} to write the entanglement entropy as a limit of correlation functions of twist operators. The symmetries of the CFT are used to evaluate those correlation functions, which lets us find the entropy. In this section, the correlation functions will be vacuum correlation functions, and we will review how the large-$c$, sparse spectrum assumption reproduces the holographic answer in these cases. In the following section, we will consider the time-dependent situation of a global quench, where the relevant correlation functions are those of a BCFT.

We calculate the entanglement entropy of a region $A$ using the replica trick, which realizes the entropy as a limit of traces of powers of the density matrix:
\be
S = \lim_{n\to 1} \frac{1}{1-n}\log {\rm Tr} \rho_A^n
\ee
A standard way to compute ${\rm Tr} \rho_A^n$ is via a path integral on an $n$-sheeted cover of the original surface, with branch points located at the endpoints of $A$. This means that the sheets of the cover are sewn together along $A$, which is the path integral representation of the matrix multiplication that defines $\rho_A^n$. Alternatively, ${\rm Tr} \rho_A^n$ can be computed as a certain correlation function in the theory ${\rm CFT}^n$, consisting of $n$ copies of the original CFT. The twist operator $\mathcal{T}_n(x)$ is defined in the ${\rm CFT}^n$ theory as the operator which implements the boundary conditions of the $n$-sheeted cover: monodromy around the twist operator shifts a local operator $\mathcal{O}_k(x)$ in the $k$th copy of the CFT to the same operator $\mathcal{O}_{k+1}(x)$ in the $k$+1st copy of the CFT. When the region $A$ consists of a union of intervals, ${\rm Tr} \rho_A^n$ can be computed (up to a constant of proportionality) as the correlation function of twist operators inserted at the endpoints of the intervals. The twist operator $\mathcal{T}_n$ is inserted at all of the left endpoints, and the antitwist operator $\mathcal{T}_{-n}$ (which sends $\mathcal{O}_k\to \mathcal{O}_{k-1}$) is inserted at the right endpoints. To actually evaluate these correlation functions, we make use of the fact that twist operators are primary with scaling dimension
\be
\Delta_n = \frac{c}{12n}(n^2-1),
\ee
where $c$ is the central charge of the CFT.

As an illustration, consider the region $A$ consisting of two intervals, $[x_1,x_2]$ and $[x_3,x_4]$, with $x_i < x_{i+1}$. Then ${\rm Tr}\rho_A^n$ can be computed as a four-point function of twist operators. Defining the cross-ratio $\eta$ as
\be
\eta = \frac{(x_2-x_1)(x_4-x_3)}{(x_3-x_1)(x_4-x_2)},
\ee
noting that $\eta\in [0,1]$, and the new coordinate $w$,
\be
w(z) = \frac{(z -  x_1)(x_4-x_3)}{(x_3-x_1)(x_4-z)},
\ee
we can write the required four-point function as
\begin{align}
\ev{\mathcal{T}_n(x_1)\mathcal{T}_{-n}(x_2)\mathcal{T}_n(x_3)\mathcal{T}_{-n}(x_4)} &=\left|\frac{\eta}{(x_2-x_1)(x_4-x_3)}\right|^{2\Delta_n}\ev{\mathcal{T}_n(0)\mathcal{T}_{-n}(\eta)\mathcal{T}_n(1)\mathcal{T}_{-n}(\infty)}\\
&=\left|\frac{1-\eta}{(x_3-x_2)(x_4-x_1)}\right|^{2\Delta_n}\ev{\mathcal{T}_n(0)\mathcal{T}_{-n}(\eta)\mathcal{T}_n(1)\mathcal{T}_{-n}(\infty)},
\end{align}
where the correlation function appearing on the right-hand side is defined via the limit
\be
\ev{\mathcal{T}_n(0)\mathcal{T}_{-n}(\eta)\mathcal{T}_n(1)\mathcal{T}_{-n}(\infty)} \equiv \lim_{w\to \infty} |w|^{2\Delta_n}\ev{\mathcal{T}_n(0)\mathcal{T}_{-n}(\eta)\mathcal{T}_n(1)\mathcal{T}_{-n}(w)}.
\ee
A general four-point function can be evaluated using the conformal block decomposition:
\be
 \ev{\mathcal{O}_1(0)\mathcal{O}_{2}(\eta)\mathcal{O}_3(1)\mathcal{O}_{4}(\infty)} = \sum_p C^p_{12}C^p_{34} \mathcal{F}(c,h_p,\{h_i\},\eta)\mathcal{F}(c,\overline{h}_p,\{\overline{h}_i\},\overline{\eta}).
\ee
The sum is over all primary operators in the theory, with $h_p$ ($\overline{h}_p$) being the (anti-) holomorphic scaling dimension of the primary operator. This sum is sometimes called the $s$-channel decomposition, but there is also a $t$-channel decomposition which features the coefficients $C^p_{23}C^p_{14}$ instead. Evaluating this sum requires an expression for the conformal blocks $\mathcal{F}$, which there are efficient algorithms for computing, as well as knowledge of the coefficients $C^p_{ij}$, which depend on the theory in question. We are only interested in correlations of twist operators, so we can set all of the $h_i$ equal to the same value $h_i= \overline{h}_i  = h = \Delta_n/2$. For small $\eta$, the Taylor series of the conformal block gives $\mathcal{F} = \eta^{h_p-2h}(1 + O(\eta))$. So at small $\eta$, the dominant term in the sum comes from $h_p=0$, which is the identity block \cite{Hartman}. This is the disconnected part of the four-point function: keeping this term alone reduces it to a product of two two-point functions:
\begin{align}
\ev{\mathcal{T}_n(x_1)\mathcal{T}_{-n}(x_2)\mathcal{T}_n(x_3)\mathcal{T}_{-n}(x_4)} &\approx \left|\frac{1}{(x_2-x_1)(x_4-x_3)}\right|^{2\Delta_n} \nonumber\\&= \ev{\mathcal{T}_n(x_1)\mathcal{T}_{-n}(x_2)}\ev{\mathcal{T}_n(x_3)\mathcal{T}_{-n}(x_4)}.
\end{align}
This is a manifestation of the cluster decomposition principle. A nontrivial fact about large-$c$ CFTs with a sparse spectrum is that this is the dominant contribution even for {\em finite} values of $\eta$, all the way to $\eta=1/2$ \cite{Hartman}. This can be proved by looking at a large-$c$ expansion of the conformal blocks, where it is seem explicitly that the identity block makes the largest contribution. Here the assumption of a sparse spectrum means that the number of operators with scaling dimensions less that $O(c)$ should not scale with $c$.

For $\eta \approx 1$, the contribution of the identity block in the $t$-channel decomposition says
\begin{align}
\ev{\mathcal{T}_n(x_1)\mathcal{T}_{-n}(x_2)\mathcal{T}_n(x_3)\mathcal{T}_{-n}(x_4)} &\approx \left|\frac{1}{(x_3-x_2)(x_4-x_1)}\right|^{2\Delta_n} \nonumber\\&= \ev{\mathcal{T}_n(x_3)\mathcal{T}_{-n}(x_2)}\ev{\mathcal{T}_n(x_4)\mathcal{T}_{-n}(x_1)}.
\end{align}
Again, at large-$c$ with a sparse spectrum this formula is expected to hold down to $\eta=1/2$. In the $c\to \infty$ limit there is a sharp phase transition between the $s$-channel and $t$-channel results at $\eta=1/2$. Finite-$c$ corrections should smooth this transition, but consideration of those effects is beyond the scope of this work. Taking the appropriate limit as $n\to 1$, these results show that the vacuum entanglement entropy of a pair of intervals at large-$c$ is given by
\be
S = {\rm min}\Big( \frac{c}{3}\log \frac{(x_2-x_1)(x_4-x_3)}{\epsilon^2}, \, \frac{c}{3}\log \frac{(x_4-x_1)(x_3-x_2)}{\epsilon^2}\Big),
\ee
where $\epsilon$ is the UV cutoff scale. This matches the holographic answer \cite{RT}.

When there are $N$ intervals, we must compute a $2N$-point function of twist operators. There are many possible decomposition channels, and the dramatic simplification at large-$c$ with a sparse spectrum is that there is always some channel in which the identity block provides the dominant contribution, and in this channel the $2N$-point function decomposes as a product of $N$ two-point functions. Taking the $n\to 1$ limit to extract the entanglement entropy, this precisely reproduces the Ryu-Takayanagi formula for arbitrary numbers of intervals \cite{Hartman, Faulkner}.

\subsection{Global Quench}

Following the setup of \cite{CC-particle-two,Cardy-quench-bcft-one,Cardy-quench-bcft-two}, the global quench is effectively modeled as a BCFT calculation. The correlation functions we need can be computed at large-$c$ with a sparse spectrum using the techniques of \cite{Hartman}. The leading-order holographic result is obtained by assuming that the only primary operator of scaling dimension less than $O(c)$ is the identity, which contributes an amount of $O(c)$ to the entanglement entropy. This will also be the dominant contribution when the spectrum is sparse, since other primary operators will contribute at $O(1)$ to the entropy. If the number of operators scales with $c$, then clearly this will compete with the identity contribution and the holographic result does not apply.

We wish to compute the four-point function of the twist operators after a quantum quench, where we take the state to be $\ket{\Psi(t)} = U(t) \ket{\Psi}$. We are interested in times well after the local equilibration time, and so it turns out to be useful to model the local equilibration by an initially Euclidean evolution over a small imaginary time $\tau_0$, which can be thought of as a regulator for the calculation. Introducing the coordinate $z= x+i\tau$, the path integral preparing the ket state at $t=\tau = 0$ is then a path integral over the strip ${\rm Im} z \in [-\tau_0, 0)$, with a boundary wavefunction at $z=-i\tau_0$, while the bra state can be obtained by integrating over the strip ${\rm Im} z \in (0,\tau_0]$. To find the correlation functions at real times $t\gg \tau_0$, we would continue to path-integrate over a real-time contour before inserting our operators, as in the Schwinger-Keldysh formalism. Alternatively, we can compute correlation functions in the strip ${\rm Im} z \in [-\tau_0, \tau_0]$ for arbitrary values of the imaginary time $\tau$ and then afterward analytically continue the answers to real time. 

For a collection of $N$ intervals, we need to compute a $2N$-point function of twist operators in the strip ${\rm Im} z \in [-\tau_0, \tau_0]$:
\be
{\rm Tr\,} \rho^n = \ev{\mathcal{T}_n(x_1+i\tau)\mathcal{T}_{-n}(x_2+i\tau)\cdots\mathcal{T}_n(x_{2N-1}+i\tau)\mathcal{T}_{-n}(x_{2N}+i\tau)}_{\rm strip},
\ee
for arbitrary values of the imaginary time $\tau\in [-\tau_0, \tau_0]$. We will take $\tau$ to be a real number initially, and then analytically continue $\tau \to it$ in the final answer to extract the real-time post-quench correlation functions. 

We can conformally transform the strip to the upper half plane by setting
\be
w(z) = \exp\left[\frac{\pi}{2\tau_0}(z+i\tau_0)\right].
\ee
Then the lines $ z = \mp i\tau_0$ map to the positive and negative real $w$ axes, respectively. The $ z = i\tau$ line maps to ${\rm arg}\, w = \pi/2 + \tau/2\tau_0$. We can write the correlation function as 
\be
{\rm Tr\,} \rho_A^n = \left|\left(\frac{\pi}{2\tau_0}\right)^{2N} e^{\frac{\pi}{2\tau_0}\sum x_i}\right|^{\Delta_n}\ev{\mathcal{T}_n\left(w_1\right)\mathcal{T}_{-n}\left(w_2\right)\cdots\mathcal{T}_n\left(w_{2N-1}\right)\mathcal{T}_{-n}\left(w_{2N}\right)}_{\rm UHP},
\ee
where the $w_i$ are the images the $x_i$.

The upper half plane correlation functions should be computed in the context of BCFT, which tells us that each primary operator in the upper half plane can be thought of as the product of a holomorphic operator at its location times another holomorphic operator at the conjugate location (reflected over the real axis) \cite{BCFT,DiFrancesco}. Then the $2N$-point function in the upper half-plane can be computed as a $4N$-point function in the full plane, which is just a vacuum correlation function. It is useful to parametrize the four-point function in terms of ${2N\choose 2}$ real parameters $\eta_{ij}$, which are invariant cross-ratios characterizing the separation of $w_i$ and $w_j$,
\be
\eta_{ij} \equiv 1 - \frac{w_{i\bar{i}}w_{j\bar{j}}}{w_{i\bar{j}}w_{j\bar{i}}} =\frac{w_{ij}w_{\bar{j}\bar{i}}}{w_{i\bar{j}}w_{j\bar{i}}},
\ee
where we have used the notation $w_{ij} = w_i - w_j$ and $w_{i\bar{j}} = w_i -\overline{w_j}$. Note that $\eta_{ij} \in [0,1]$. Also, the UHP $2N$-point function we started with should only depend on $4N-3$ real degrees of freedom (after making use of the part of the conformal symmetry which maps the real axis to itself), so the $\eta_{ij}$ parameters are not all independent. They are still a useful parametrization, however. When $\eta_{ij} \approx 1$, the operators at $w_i$ and $w_j$ are much closer to their respective image points than to each other. Likewise, at $\eta_{ij} \approx 0$ the operators at $w_i$ and $w_j$ are closer to each other than to their respective images. This behavior helps determine efficient OPE expansion channels, as we will see below.

The time-dependence of the correlation function is reflected in the time-dependence of the $\eta_{ij}$ cross-ratios, which in the $\tau_0\to 0$ limit are given by
\be\label{eq-etatimedep}
\eta_{ij}(t) = 1-\frac{2\cosh^2 (\pi t/2\tau_0)}{\cosh(\pi |x_i - x_j|/2\tau_0)+\cosh(\pi t/\tau_0)} \approx \frac{1}{1+\exp\left[\frac{\pi}{\tau_0}\left(t-\frac{|x_i-x_j|}{2}\right)\right]}.
\ee
We see that there are sharp transitions between $\eta_{ij} \approx 1$ and $\eta_{ij} \approx 0$, which occur at half the light-crossing time, $|x_i-x_j|/2$.

At early times (meaning for times less than the length scales defining the intervals, but still much greater than the local equilibrium scale $\tau_0$), all of the $\eta_{ij}$ are approximately equal to one. Then the upper half-plane $2N$-point function approximately factorizes into $2N$ full-plane two-point functions: this is the cluster decomposition limit where each operator in the upper half plane is paired with its image point in the lower half plane. Then we find
\be
{\rm Tr\,} \rho^n =  \left|\frac{\pi}{2\tau_0}e^{-\pi t/2\tau_0} \right|^{2N\Delta_n}.
\ee
This leads to a linear growth of the entanglement entropy at a rate which is $N$ times as fast as it would have been for a single interval (in agreement with the quasiparticle picture).

At very late times, all of the $\eta_{ij}$ parameters are very small. One can check that this corresponds to a different cluster decomposition limit: now adjacent operators in the upper half-plane are paired with each other, and their image points are paired with each other. So again we find a product of $2N$ full-plane two-point functions. One can check that in this limit the thermal entropy formula $S = s_{\rm eq}V_A$ is produced.

At intermediate times, when some of the $\eta_{ij}$ are small and others are approximately equal to one, we need to be more careful. This is where the real difference between holographic and weakly coupled CFTs lies. To simplify the notation, we will restrict ourselves to $N=1$ and $N=2$, though similar arguments hold for all $N$.

At $N=1$ there are no real surprises, but it is useful to go through it to illustrate some of the key points. There is only a single interval and a single cross-ratio $\eta$, and the two-point function in the upper half-plane can be written as a four-point function in the full plane, which we analyzed above. The conformal block decomposition of the four-point function implies that we can write the UHP two-point function as
\be
\ev{\mathcal{T}_n\left(w_1\right)\mathcal{T}_{-n}\left(w_2\right)}_{\rm UHP} = \frac{1}{|w_{1\bar{1}}w_{2\bar{2}}\eta|^{2\Delta_n}}F_n(\eta).
\ee
Even though $F_n(\eta)$ may be a very complicated function, the cluster decomposition limits tell us that $F_n \to 1$ both when $\eta\to 0$ and $\eta\to 1$, since the prefactor alone reproduces both of those limits. This makes the global quench computation easy to perform when $\tau_0\to 0$ (i.e., at times and distances much greater than the equilibration scale), since $\eta$ makes a rapid transition from zero to one in that case. Then for arbitrary CFTs, the factor $F_n(\eta)$ is completely inconsequential and can be dropped; the result agrees with both the quasiparticle model and the holographic calculation. However, we would like to point out that a stronger statement can be made about large-$c$ CFTs with a sparse spectrum. In the limit $c\to \infty$, we have an exact formula for $F_n(\eta)$ that comes from demanding that the full-plane four-point function always factorize in either the $s$-channel or $t$-channel:
\be
F_n(\eta) = \begin{cases}
\eta^{\Delta_n},& ~~~\eta > 1/2,\\
(1-\eta)^{\Delta_n},& ~~~\eta < 1/2.
\end{cases}
\ee
Although this formula did not make a difference for a single interval, we will now see how things change with multiple intervals.

With two intervals, $N=2$, the relevant correlation function is a four-point function in the upper half-plane. Following \cite{Coser:2014gsa}, we will parametrize the four-point function as
\be\label{eq-lowcscaling}
\ev{\mathcal{T}_n\left(w_1\right)\mathcal{T}_{-n}\left(w_2\right)\mathcal{T}_n\left(w_3\right)\mathcal{T}_{-n}\left(w_4\right)}_{\rm UHP} = \frac{1}{\prod_i |w_i - \bar{w}_i|^{\Delta_n}}\left(\frac{\eta_{13}\eta_{24}}{\eta_{12}\eta_{23}\eta_{14}\eta_{34}}\right)^{\Delta_n} F_n(\left\{\eta_{ij}\right\}).
\ee
The function $F_n$ (different from above, but with the same name) is in principle very complicated, but since we only care about $\eta_{ij} \approx 1$ or $\eta_{ij} \approx 0$, only the values of $F_n$ at those points are necessary to find the entanglement entropy. So as time passes and the $\eta_{ij}$ transition from one to zero, $F_n$ effectively becomes a piecewise constant function of time. This translates into an additive term in the entropy, which is usually argued to be subleading in the $\tau_0\to 0$ limit. However, this reasoning does not apply if the leading constant term in $F_n$ is zero during any given phase. Then we have to consider terms which are proportional to some of the nearly-vanishing $\eta_{ij}$, which are exponentially small and in particular time-dependent. This is what happens for theories with a holographic dual. At large-$c$ and with a sparse spectrum, the eight-point function in the full plane, which is equal to the four-point function we wish to compute in the upper half-plane, always factorizes in some channel into a product of two-point functions~\cite{Hartman}. By comparing this result to \eqref{eq-lowcscaling}, we find that the effective $F_n$ factors are proportional to the $\eta_{ij}$ whenever the quasiparticle model (i.e., the prefactor in \eqref{eq-lowcscaling}) says that the entropy should decrease. In particular, the resulting time dependence of $F_n$ precisely cancels out the time dependence of the prefactor, meaning that the entropy remains saturated at its thermal value and does not decrease.

\section{Entanglement Tsunami}\label{sec-tsunami}

It would be desirable to have a picture of propagating interacting quasiparticles to replace the free-streaming quasiparticle picture that seems to work well for free theories. Such a picture should reproduce the linear growth rate evident at early times, but avoid decreases in entropy. We have not been able to derive such a picture in terms of particles, but a simple heuristic which gives the correct answer is the ``entanglement tsunami" of \cite{HL-one,HL-two}, which we will elaborate on here. We will begin with a discussion that applies to one or two disjoint intervals, where the holographic entropy can be reproduced exactly, and then explain how to extend the idea to arbitrary numbers of disjoint intervals where we only have an upper bound on the entropy.

\subsection{One or Two Intervals}

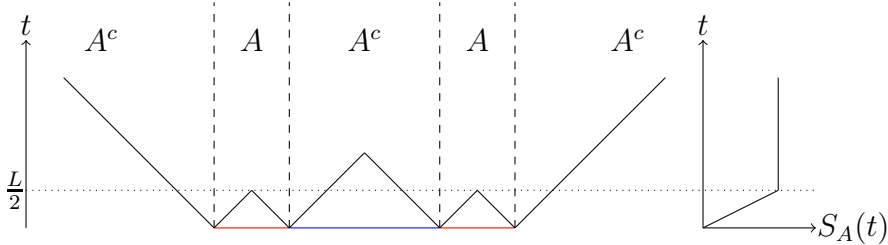
\begin{figure}
\centering 
\begin{tikzpicture}[baseline]

\draw[red] (-2,0) --(-1,0);
\draw[red] (1,0) --(2.0,0);
\draw[blue] (-1,0) --(1,0);
\draw (-2,0) --(-4,2);
\draw (-2,0) --(-1.5,.5);
\draw (-1.5,.5) --(-1,0);
\draw (-1,0) --(0,1);
\draw (0,1) --(1,0);
\draw (1,0) --(1.5,.5);
\draw (1.5,0.5) --(2,0);
\draw (2,0) --(4,2);

\draw[dashed] (-2,0) --(-2,3);
\draw[dashed] (-1,0) --(-1,3);
\draw[dashed] (1,0) --(1,3);
\draw[dashed] (2,0) --(2,3);
		\node at (-3.5,2.5) {$A^{c}$};
		\node at (-1.5,2.5) {$A$};
		\node at (0,2.5) {$A^{c}$};
		\node at (1.5,2.5) {$A$};
		\node at (3.5,2.5) {$A^{c}$};
	
	\node at (4.5,2.7) {$t$};
	\node at (6.5,0.0) {$S_{A}(t)$};

	\draw[->] (-4.5,0) --(-4.5,2.5);
	\node at (-4.5,2.7) {$t$};
	\node at (-4.65,0.5) {$\frac{L}{2}$};
	
	\draw[->] (4.5,0) --(4.5,2.5);
	\draw[->] (4.5,0) --(6.0,0);
	\draw (4.5,0) --(5.5,0.5);
	\draw (5.5,.5) --(5.5,2.0);
	
	\draw[dotted] (-4.5,0.5) --(6,0.5);

		\end{tikzpicture}%

\caption{The quench for two intervals of length $L$ separated by a distance $R$ when $L >R$. On the left, we show the entanglement tsunami wavefront as a function of time (jagged black line.) The region $A$ is marked as red. The intervals between the disconnected components of $A$ are marked as blue. On the right we show the entanglement entropy as a function of time.}\label{fig-simpletsunami}	
		\end{figure}

Our picture of the entanglement tsunami is as a wave which begins at each of the endpoints of $A$ at $t=0$ and flows outward in both directions (see Fig.~\ref{fig-simpletsunami}). This wave is not a physical wave representing the propagation of particles or energy: it is merely a tool for understanding the entanglement. In particular, note that the quench state is homogeneous, while the wave begins at particular locations picked out by the region $A$ we have chosen. At any time $t>0$, the wave divides the space into two regions: one which has already been overtaken by the wave, and one which has yet to be overtaken by the wave. We will call the former region the ``entangled" region, and the latter region the ``unentangled" region. This picture is reminiscent of a vacuum decay, where at $t=0$ we have tunneling events at each of the endpoints of $A$ and the entanglement tsunami wavefront is like a bubble wall which converts the metastable vacuum (the unentangled region) into the true vacuum (the entangled region). (Though we should emphasize once more that there is no physical sense in which the entanglement tsunami is changing the vacuum; it just has the same cartoon picture.) If $A$ has more than one endpoint, i.e., it is not just a half-line, then eventually the entanglement tsunami wavefronts starting from adjacent endpoints will collide. For all times after the collision, the entire interval between those two endpoints will be part of the entangled region.

We still have to give a rule for computing the entanglement entropy of the region $A$ given this tsunami picture. Let the entangled region at time $t$ be denoted by $E(t)$. When $A$ consists of just one or two intervals, then its entanglement entropy is
\be\label{eq-simpletsunami}
S(t) = s_{\rm eq} \times {\rm min}\Big({\rm Vol}(E(t)\cap A), {\rm Vol}(E(t)\cap A^c) \Big).
\ee
It is not hard to see that this rule agrees with the holographic prescription for entanglement entropy: for one interval it is trivial, and for two intervals the two options essentially coincide with the two possible HRT surfaces. Also, note that the entropy is symmetric with respect to $A$ and $A^c$, which is as it should be for a globally pure state.

Although we have emphasized that the entanglement tsunami does not represent the propagation of a physical excitation, there is a suggestive interpretation in which we do imagine a collection of excited particles living in $E(t)$. Suppose that a finite density of qubits populates the region $E(t)$, and that those quibits are in a typical pure state. Then the entanglement entropy of the qubits in $E(t)\cap A$ will follow the Page rule, meaning that their entanglement entropy will be proportional to either the number of qubits in $E(t)\cap A$ or $E(t)\cap A^c$, whichever is smaller \cite{Page}. This is precisely the entanglement tsunami prescription.
 
 \subsection{Multiple Intervals}
 
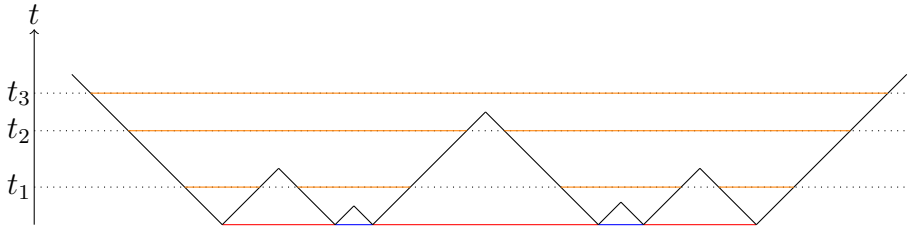
\begin{figure}
\centering 
\begin{tikzpicture}[baseline]

\draw[red] (-2,0) --(-0.5,0);
\draw[red] (0,0) --(3,0);
\draw[red] (3.6,0) --(5.1,0);
\draw[blue] (-.5,0) --(0,0);
\draw[blue] (3,0) --(3.6,0);

\draw (-2,0) --(-4,2);
\draw (-2,0) --(-1.25,.75);
\draw (-.5,0) --(-1.25,.75);
\draw (-.5,0) --(-0.25,.25);
\draw (0,0) --(-0.25,.25);
\draw (0,0) --(1.5,1.5);
\draw (3,0) --(1.5,1.5);
\draw (3,0) --(3.3,0.3);
\draw (3.6,0) --(3.3,0.3);
\draw (3.6,0) --(4.35,0.75);
\draw (5.1,0) --(4.35,0.75);
\draw (5.1,0) --(7.1,2.0);




\draw[dotted] (-4.5,0.5) --(7.1,.5);
\draw[dotted] (-4.5,1.75) --(7.1,1.75);
\draw[dotted] (-4.5,1.25) --(7.1,1.25);
\draw[orange] (-2.5,0.5) --(-1.5,0.5);
\draw[orange] (0.5,0.5) --(-1,0.5);
\draw[orange] (4.1,0.5) --(2.5,0.5);
\draw[orange] (4.6,0.5) --(5.6,0.5);

\draw[orange] (-3.25,1.25) --(1.25,1.25);
\draw[orange] (1.75,1.25) --(6.35,1.25);

\draw[orange] (-3.75,1.75) --(6.85,1.75);

\draw[->] (-4.5,0) --(-4.5,2.6);
\node at (-4.5,2.8) {$t$};
\node at (-4.7,0.5) {$t_{1}$};
\node at (-4.7,1.75) {$t_{3}$};
\node at (-4.7,1.25) {$t_{2}$};

		\end{tikzpicture}%

\caption{Entanglement tsunami for many intervals. The region $A$ is marked as red. The intervals between the disconnected components of $A$ are marked as blue. Note that at time $t_{1}$, $E(t)$ consists of four disconnected components (orange solid lines) separated by the entanglement tsunami wavefront (jagged black line). But at time $t_{2}$, the first pair and second pair have merged, leaving two disconnected regions. At time $t_{3}$, there is only a single connected region.}\label{fig-multitsunami}	
		\end{figure}

 When $A$ consists of more than two intervals, there does not appear to be a simple rule like \eqref{eq-simpletsunami} which correctly reproduces the holographic answer, but right answer. One can attempt the following simple and natural generalization, which involves a refinement of our notion of the entangled region $E(t)$. Instead of a single region, the entangled region is naturally the union of disjoint intervals:
 \be
 E(t) = \bigcup_i E_i(t),
 \ee
 where each $E_i(t)$ is an interval representing a single connected component of $E(t)$. The number of $E_i(t)$ regions changes with time. For $N$ intervals, $E(t)$ has $N+1$ connected components at early times (each being a small neighborhood around an endpoint of an interval) and only a single connected component at late times. A collision of entanglement tsunami wavefronts indicates that two of the $E_i(t)$ are merging, and thereafter will be treated as a single unit. We illustrate this behavior in Fig.~\ref{fig-multitsunami}. 
 
 With this refinement of the entangled region, we have the following upper bound on the holographic entanglement entropy\footnote{We would like to thank Mark Mezei for providing us with an example of three disjoint intervals where equality fails.}:
 \be\label{eq-multitsunami}
 S(t) \leq s_{\rm eq} \sum_i {\rm min}\Big({\rm Vol}(E_i(t)\cap A), {\rm Vol}(E_i(t)\cap A^c) \Big).
 \ee
In other words, we have a separate minimization problem for each connected component of the entangled region, and at the end we add them all up. This rule is numerically the same as \eqref{eq-simpletsunami} for one or two intervals, and so there is equality. For more than two intervals, \eqref{eq-multitsunami} corresponds to the area of one of the candidate HRT surfaces, though not necessarily the minimal one. Therefore it represents only an upper bound on the entropy. Forthcoming work from Casini, Liu, and Mezei discusses such an upper bound beyond the context of holography~\cite{CasiniLiuMezei}. We also note that \eqref{eq-multitsunami} is manifestly symmetric with respect to $A$ and $A^c$, even though each minimization subproblem is free to choose to use region $A$ or $A^c$ independently for its contribution to the entropy.

 There is also still a suggestive qubit picture for the right-hand side of \eqref{eq-multitsunami}. Now instead of one collection of qubits in $E(t)$, there is an independent collection of qubits in each of the $E_i(t)$, and the total qubit state is a product state over the connected components. We can imagine that the qubits are somehow being emitted at the endpoints of the intervals, and when two qubit chains come into contact (i.e., when two wavefronts merge), the two qubit chains undergo a very rapid mixing and end up in a typical pure state of the composite system. Even though this picture only provides an upper bound on the entropy, it could be that a similar picture is accurate for the real system.

\section{Discussion}\label{sec-disc}

\subsection{Summary}

We have seen that for a (1+1)-dimensional CFT in the limit of large-$c$, with a sparse spectrum of operators, the time-dependence for the entanglement entropy of multiple intervals is not correctly captured by the quasiparticle model of \cite{CC-particle,CC-particle-two}. The correct answer, which is correctly reproduced holographically, is non-decreasing with time, whereas the quasiparticle model features crops in the entropy. A heuristic model which gives the correct time evolution is provided by the entanglement tsunami, a picture originally introduced in \cite{HL-one, HL-two} which we have developed into a rule for calculating the entanglement. There are many unanswered questions about this model, which we will now discuss.

\subsection{Higher Dimensions}

In higher dimensions, as in (1+1)-dimensions, the entanglement entropy for disjoint regions in the quasiparticle picture will experience periods of decrease that are not present in the holographic calculation. The entanglement tsunami picture will solve this problem since, like the holographic calculation, it does not allow for decreases in the entropy. There remains the puzzle of the velocity of the entanglement tsunami wavefront. In (1+1)-dimensions, it seems natural to say that the wavefront moves at the speed of light. But this would cause the entanglement to grow more quickly than the holographic calculation indicates. At the same time, a wavefront made up of quasiparticles traveling in random directions moves too slowly. A field-theoretic derivation of the entanglement tsunami velocity would do much to clarify the picture.

\subsection{Interacting Quasiparticles?}

A natural guess for fixing up the quasiparticle picture is simply to add interactions. Instead of free-streaming, the particles would bump into each other and generate multipartite entanglement. It is not clear whether a simple picture of this type can accurately reproduce the holographic calculation, but the entanglement tsunami prescription may provide a useful starting point. In the entanglement tsunami, it appears as though there are clouds of particle being emitted by the endpoints of the the region $A$ (or, in higher dimensions, by the boundary of $A$), beginning at the quench time. The unphysical aspect of this model is that it is not homogeneous: the boundaries of $A$ are not preferred locations in the state, so it doesn't make sense to say that they are the sources of particles. Instead, the state should describe a finite density of interacting particles everywhere at once. However, since we are asking about the entanglement entropy of $A$, there may be a sense in which the particles in a neighborhood of the boundary of $A$ are the only ones that matter. Then the entanglement tsunami would represent a growing sphere of influence, picking out the parts of the homogenous matter distribution which determine the entanglement. We leave an exploration of this possibility to future work.

\acknowledgments

We would like to thank C.~Akers, R.~Bousso, T.~Faulkner, Z.~Fisher, T.~Grover, V.~Hubeny, A.~Kitaev, J.~Koeller, M.~Mezei, R.~Myers, S.~Shenker, M.~Smolkin, D.~Stanford, J.~Suh, and C.~Zukowski for useful discussions and correspondence, as well as all of the participants in the ``Quantum Gravity Foundations: UV to IR" and ``Entanglement in Strongly-Correlated Quantum Matter" programs at KITP, Santa Barbara. This work was supported in part by the Berkeley Center for Theoretical Physics, by the National Science Foundation (award numbers 1214644 and 1316783), by fqxi grant RFP3-1323, and by the US Department of Energy under Contract DE-AC02-05CH11231.

\bibliographystyle{utcaps}
\bibliography{ent}

\end{document}